\documentclass{iopart}
\usepackage{graphicx} 
\usepackage{rotating}
\usepackage{cite}
\usepackage{float}

\begin{document}

\title{Systematical, experimental investigations on LiMg$Z$ ($Z$= P, As, Sb) wide band gap semiconductors}

\author{Andreea Beleanu, Mihail Mondeshki, Quin Juan, Frederick Casper, Claudia Felser}
\address{Institut f\"ur Anorganische Chemie und Analytische Chemie, Johannes Gutenberg-Universit\"at Mainz, Staudinger Weg 9, 55128 Mainz, Germany}
\ead{felser@uni-mainz.de}
\author{Florence Porcher}
\address{Laboratoire Leon Brillouin, CEA Saclay, 91191 Gif-sur-Yvette Cedex, France}

\date{\today}

\begin{abstract}

This work reports on the experimental investigation of the wide band gap compounds LiMg$Z$ ($Z$=~P, As, Sb), which are promising candidates for opto-electronics and anode materials for Lithium batteries. The compounds crystallize in the cubic ($C1_b$) MgAgAs structure (space group $F\bar{4}3m$). The polycrystalline samples were synthesized by solid state reaction methods. X-ray and neutron diffraction measurements show a homogeneous, single-phased samples. The electronic properties were studied using the direct current (DC) method. Additionally UV-VIS diffuse reflectance spectra were recorded in order to investigate the band gap nature. The measurements show that all compounds exhibit semiconducting behavior with direct band gaps of 1.0 eV to 2.3 eV depending on the $Z$ element. A decrease of the peak widths in the static $^7$Li nuclear magnetic resonance (NMR) spectra with increasing temperature was observed, which can directly be related to an increase of Li ion mobility.

\end{abstract}



\submitto{\JPD}

\maketitle

\section{Introduction}
The class of semiconductors commonly referred to as "wide band gap semiconductors" hold promise for continued revolutionary improvements in the size, cost, efficiency and performance of a broad range of microelectronic and optoelectronic application~\cite{scribd}. Binary semiconductors can be realized from wide to low band gap (ZnO to HgTe). However very often the compounds show intrinsic defects~\cite{VRF11,RM05} or indirect band gaps~\cite{KR78}, which makes a search for new semiconducting compounds necessary. However ternary semiconductors were not investigated systematically up to date. This contribution focus on the ternary semiconducting compounds LiMg$Z$ ($Z=$ P, As, Sb). The band gaps have been predicted theoretically~\cite{CZW85} to be \textit{direct} with a band gap of 2.43~eV for LiMgP~\cite{KK00,KKNFG10,KKT88}, 2.31~eV for LiMgAs~\cite{KBM06,KYS10} and 2.0~eV for LiMgSb~\cite{G10}.

The compounds LiMg$Z$ ($Z=$ P, As, Sb) belong to the so called Nowotny-Juza compounds. Nowotny-Juza compounds A$^{I}$B$^{II}$C$^{V}$ based on three main group elements (A, B and C) exhibit the same crystalline structure ($C1_b$) as the \textit{XYZ} Heusler compounds, where \textit{X} and \textit{Y} are transitions metals and \textit{Z} is a main group element. They can be viewed as zinc-blende III-V compounds in which the III column has been "disproportionated" into A$^{I}$+B$^{II}$ atoms~\cite{BC88}.

The wide band gaps of the compounds make them  promising candidates for opto-electronics, ranging from blue lasers to cadmium free solar cell materials (substituting CdS, CdSe, CdTe) and buffer layer materials for chalcopyrite-based thin film solar cell devices~\cite{GFP11,KKNFG10,G10}. The structure of the compounds was determined by x-ray and neutron diffraction. Nuclear magnetic resonance (NMR) spectroscopy was used to investigate the Li ion mobility . The band gaps were measured by optical reflectance spectrometry and electrical conductivity measurements.

\section{Experimental Details}

The ternary compounds LiMg$Z$ ($Z$=~P, As, Sb) were synthesized by solid state reaction using stoichiometric amounts of Lithium ingots, Magnesium wire, Phosphorus, Arsenic and Antimony powder (all 99~\% purity, Sigma Aldrich). A tantalum crucible sealed by arc-melting under argon atmosphere was used as container for the preparation of LiMgSb. For LiMgP and LiMgAs an alumina crucible was used. The charged crucibles were sealed in quartz tubes at $10^{-3}$~mbar. Colored powders, yellow for LiMgP, yellow-brownish for LiMgAs and black for LiMgSb were obtained after heating the samples at 900$^\circ$C for 12h in a muffle type furnace. X-ray powder diffraction measurements were performed using a Siemens D5000 diffractometer with Cu-K$_\alpha$ radiation, $\lambda$=1.5418~{\AA}, in Bragg geometry. Powder neutron diffraction measurements were performed at Laboratoire Leon Brillouin (LLB) (Saclay, France) on the high resolution powder diffractometer 3$T{_2}$ using a Ge (335) monochromator to select a wavelenght of 1.2253~{\AA}. The samples were incapsulated in vanadium crucible. The experimental x-ray and neutron diffraction patterns were refined using the FullProf suite of programs~\cite{R93}.

The electrical transport was measured by the direct current (DC) method using pellets of round shape (diameter: 13~mm; thickness 1~mm) obtained by cold pressing using a hydraulic press (Perkin-Elmer) under a pressure of ca.10~bar. The electric properties of the samples were obtained by impedance spectroscopy (IS) measurements. Li conductivity, $\sigma$ DC, were determined from the DC plateau of the corresponding impedance spectra. IS was recorded as a function of temperature from -120$^\circ$~C to 180$^\circ$~C [Solartrin SI 1260 impedance gain phase analyzer with a high-resolution dielectric converter-Alpha high-resolution dielectric analyzer Novocontrol~\cite{BMW07} in the range from $10^{-1}$ - $10^{6}$~Hz]. The measurements were performed using stainless steel or platinum electrodes with 10 or 20~mm diameter under nitrogen flow. 

The variable temperature (VT) static solid state NMR measurements were conducted using a 400~MHz Bruker Avance spectrometer at a frequency of 155.40~MHz and 8~kHz spinning at the magic angle for $^7$Li. The measurements were recorded in the temperature range of 237~K to 390~K in 10 degree steps static and at ambient temperature. 5$\mu$s pulse length for the 90$^\circ$ single pulse excitation experiment and recycle delays of 120~s were used in all cases. 8 transients were averaged when acquiring the $^7$Li NMR signal. A commercial 3-channels Bruker 4 mm probe-head, capable of fast MAS was used for all measurements. The spectra were referenced to external 3M LiCl water solution.

The UV-Vis diffuse reflectance spectra were acquired on a Varian UV-VIS-NiR Cary 5G spectrophotometer equipped with an integrated sphere (Ulbricht sphere) over the 200-800 nm. The measurements were performed at room temperature. A PTFE (polytetrafluoroethylene) plate was used as reference material.

\section{Results and Discussion}

\subsection{Structure refinement}

\begin{figure}[H]
\centering
\includegraphics[width=10cm]{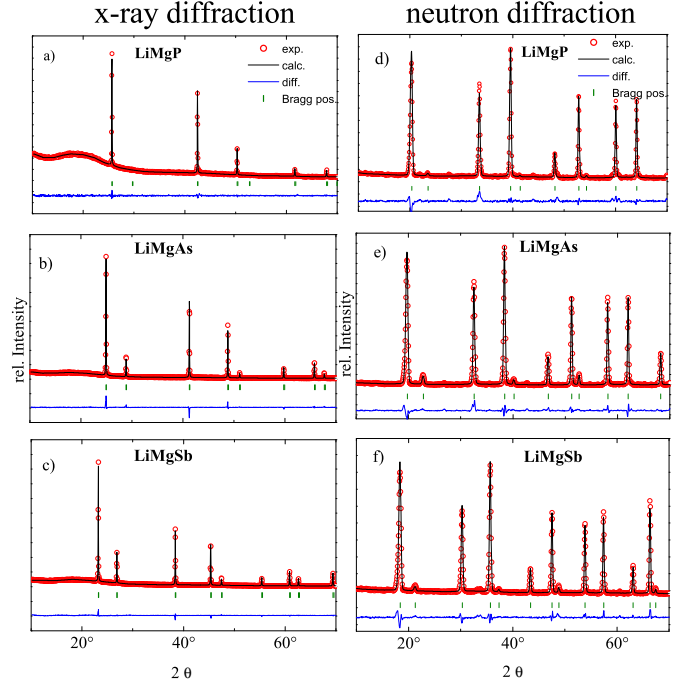}
\caption{Rietveld refinement of powder x-ray ( a, b, c) and neutron diffraction pattern ( d, e, f) of  LiMg$Z$ ($Z$=~P, As, Sb)}
\label{fig:1neutrx}
\end{figure}

X-ray diffraction on LiMg$Z$ ($Z$= P, As,Sb) showed the single homogeneous phase. Neutron diffraction measurements show additional peaks suggesting a distortion of the structure that is not observed by x-ray diffraction (Figure~\ref{fig:1neutrx}).
The Rietveld refinement on the neutron diffraction patterns confirmed the antifluorite structure. The Li atoms occupy the 4b Wickoff position and Mg and P the 4a and 4c position, respectively. The lattice parameters are summarized in Table~\ref{tab:trans}. Figure~\ref{fig:1neutrx} a), b) and c) show the measured powder diffraction pattern of the polycrystalline LiMg$Z$ ($Z$=~P, As, Sb) samples and c), d) and e) the measured neutron diffraction pattern collected at room temperature. The figures represent the fit of the diffractograms (red: observed, black: calculated by Rietveld and blue: difference plot). The lattice constants of LiMg$Z$ ($Z$=~P, As, Sb) increase from LiMgP to LiMgSb as expected with the increase of the radii of the corresponding pnictides. The lattice parameters are in good agreement with previously reported ones (~\cite{KK00,KKNFG10,KKT88,KBM06,KYS10}).

\begin{table}[htb]
\centering
\caption{Lattice constants \textit{a} (in {\AA}) obtained by Rietveld refinement on x-ray and neutron diffraction pattern of LiMg$Z$ ($Z$=~P, As, Sb)}
  \begin{tabular}{lcccc}
      method                     &   x-ray diffraction &   neutron diffraction  \\
      \hline
   LiMgP   &  6.010[9]  &  5.998[6]  \\
   LiMgAs           &  6.183[5]  &  6.176[1]  \\
   LiMgSb              &  6.628[5]  &  6.622[6]  \\
  \end{tabular}
\label{tab:trans}
  \end{table}

\subsection{UV-VIS}

For the study of the optical properties of the samples, the band gap nature was determined using a UV-VIS-NiR spectrophotometer in diffuse reflectance at room temperature over a range from 200 nm to 800 nm (6.20~eV to 1.55~eV). The optical absorption spectra of powder samples were converted from diffuse reflectance spectra using the Kubelka-Munk function \cite{KM31}:

\begin{equation}
F(R_\infty)=\frac{K}{S}=\frac{({1-R_\infty})^{2}}{2R_\infty},
\end{equation}

where $R_\infty$ is the diffuse reflectance for an infinitely thick sample, $K$ and $S$ are the Kubelka-Munk absorption and scattering coefficients, respectively. In most cases, $S$ can be considered as a constant independent of wavelength. In the parabolic band structure the band gap ${E_g}$ and absorption coefficient are related through the well-known Tauc equation \cite{S06}:

\begin{equation}
\alpha h \nu={C_1}{(h \nu-{E_g})^\gamma},
\end{equation}

where $\alpha$ is the linear absorption coefficient of the material, $h \nu$ is the photon energy, ${C_1}$ is a proportionality constant and $\gamma$ is constant depending on the band-gap nature: $\gamma$=1/2 for allowed direct band gap and $\gamma$=2 for indirect band gap. Assuming the material scatters in perfectly diffuse manner, the $K$-$M$ absorption coefficient $K$ becomes equal to $2 \alpha$. In this case, the Tauc equation becomes:

\begin{equation}
F(R_\infty) h \nu={C_2}{(h \nu-{E_g})^\gamma},
\end{equation}

Through the Tauc plot $[F (R_\infty) h \nu)]^{1/\gamma}$ the band gap ${E_g}$ of a powder sample is extracted by linear extrapolating to the energy axis.

\begin{figure}[H]
\centering
\includegraphics[width=8cm]{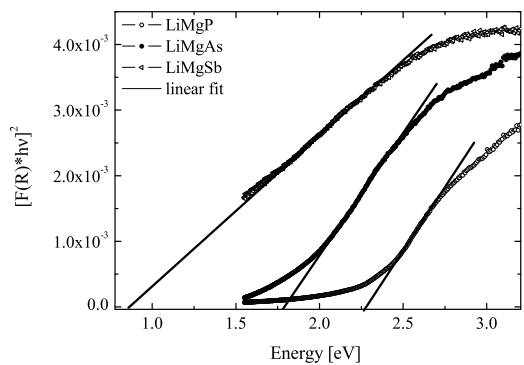}
\caption{The Tauc plots for LiMg$Z$ ($Z$= P, As, Sb) as direct semiconductors. The lines mark the linear part used to estimate the band gap.}
\label{fig:2VISb}
\end{figure}

Figure~\ref{fig:2VISb} shows the Tauc plots for samples as direct band gap semiconductors. The energy gaps obtained are summarized in Table \ref{tab:trans2}. They are 2.3 eV for LiMgP, 1.8 eV for LiMgAs and 0.9 eV for LiMgSb. The plots as indirect semiconductors (not presented) contain linear parts only in a small photon energy range. This leads to the conclusion that the converted absorption data obeys the relation of directness, which is in accord with previous reports~\cite{KKT88,KBM06,KYS10}. 

\subsection{Nuclear Magnetic Resonance Spectroscopy (NMR) measurements}

\begin{figure}[H]
\centering
\includegraphics[width=8cm]{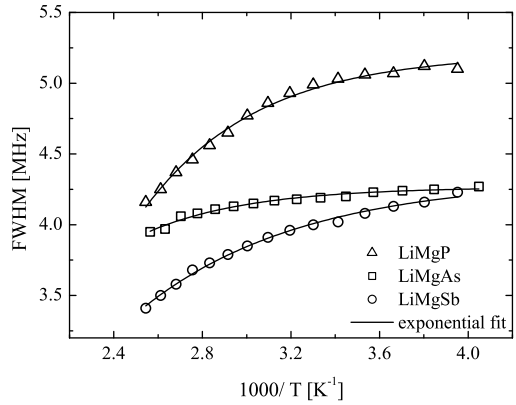}
\caption{The full width at half maximum of LiMg$Z$ ($Z$=~P, As, Sb) measured as a function of temperature. The lines mark the exponential fits .}
\label{fig:3NMR}
\end{figure}

Lineshape analysis of the $^7$Li static resonances recorded as a function of temperature was applied to investigate the thermal behavior of the compounds. The full width at half maximum (FWHM) of the $^7$Li signals in all compounds was determined in the temperature range 237-293~K and correlated with the long range Li ion transport, as measured by impedance spectroscopy. 
It is well-known that in the rigid lattice regime the random distribution of lithium ions over the surplus of available sites, as well as the presence of structural defects, results in a distribution of the electric field gradients. Thus, the quadrupolar pattern, typical of a nucleus of spin 3/2 looks smeared, with badly resolved satellites at the foot of the peak of the central transition. At higher temperatures above the onset of motional narrowing, the satellites disappear completely with only the single line of the central transition remaining. Further heating often results in a reappearance of the quadrupolar spectrum, however, with a smaller width ~\cite{BM98}.
As the central line is influenced predominantly by dipole-dipole interactions ~\cite{BJV07}, magic angle spinning at 8~kHz was applied for the LiMg$Z$ series to determine the signal maxima in separate experiments conducted at ambient conditions (supporting information). 
$^7$Li chemical shifts of 3.65~ppm, 3.10~ppm and 0.78~ppm were measured for the P, As and Sb analogues, respectively. The observed shift to higher fields is directly related with the changes in the electronic environment of the respective lithium nucleus which experiences higher shielding increasing the atomic number of the neighboring $Z$ element.
Figure~\ref{fig:3NMR} presents the FWHM of the $^7$Li NMR resonances in the static spectra of LiMg$Z$ ($Z$=~P, As, Sb) analogues with the respective exponential fits. The clear tendency of decreasing peak widths with increasing temperature is directly related to the increased Li ion mobility, that leads to motional narrowing. A correlation time of the Li ion hopping motion could not be extracted, because in the present experiments lower temperatures were not accessible to observe the onset of the narrowing.

\subsection{DC Conductivity}
\begin{figure}[H]
\centering
\includegraphics[width=8cm]{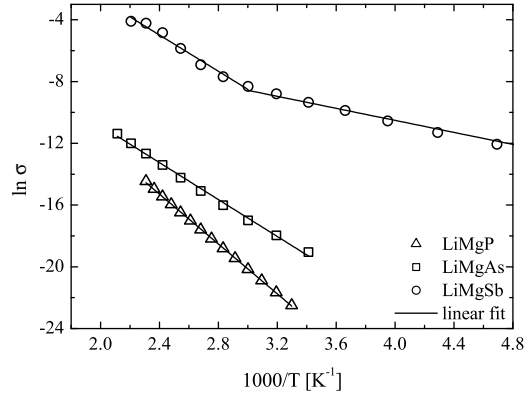}
\caption{Temperature dependence of the DC conductivity.}
\label{fig:4con}
\end{figure}

The temperature dependence of the DC electrical conductivity for LiMg$Z$ ($Z$=~P, As, Sb) from 153 to 473 K is shown in Fig.\ref{fig:4con} in which the lines correspond to linear fits to the data. The data were analyzed by an Arrhenius plot according to:
\begin{equation}
\sigma_{dc}\propto\exp(-\frac{E{_a}}{k{_B}T}), 
\end{equation}

where ${k{_B}}$ is Boltzmann's constant, ${T}$ is the temperature and ${E{_a}}$ is the activation energy. For intrinsic semiconductors the activation energy ${E_{a}}$ has a value about half of the band gap (${E_{a}}$=${E_g}$/2)~\cite{K73}. The activation energy ${E{_a}}$ is determined by the Arrhenius plot of this expression, as listed in Table \ref{tab:trans2}. From Fig.\ref{fig:4con} it is clear that with the increase of atomic number of the cation from ${P}$ to ${Sb}$, the samples show higher conductivity. Near room temperature LiMgSb shows the highest electrical conductivity of 1.5$\cdot$$10^{-4}$~S/cm while LiMgP shows the lowest of 3.2$\cdot$$10^{-10}$~S/cm. Within the measured temperature range, LiMgP and LiMgAs exhibit intrinsic conducting behavior with ${E{_a}}$ = 0.74 eV and 0.52 eV, respectively. On the contrary, LiMgSb shows two sections of activated conducting with a small ${E{_a}}$ of 0.19 eV indicating doped levels at low temperatures and a larger ${E{_a}}$ of 0.48 eV at high temperatures. The latter value is close to ${E_g}$/2 for LiMgSb obtained in Fig.\ref{fig:2VISb}, suggesting intrinsic conducting behavior at this temperature range.

\begin{table}[htb]
\centering
\caption{Electrical conductivity $\sigma$ , activation energy ${E{_a}}$ and optical band gap energy of LiMg$Z$ ($Z$=~P, As, Sb)}
  \begin{tabular}{lcccc}
      $Z$                     &  P &  As &  Sb  \\
      \hline
   $\sigma$ [S/cm]   &  3.23$\cdot$$10^{-10}$  &  1.58$\cdot$$10^{-8}$  &  1.51$\cdot$$10^{-4}$  \\
   ${E{_a}}$ [eV]           &  0.73  &  0.52    &  0.48  \\
   ${E_g}$ [eV]             &  $\approx$2.30  &  $\approx$1.80    &  $\approx$0.9  \\
  \end{tabular}
\label{tab:trans2}
\end{table}

\subsection{Summary}

In summary it has been showed that the LiMg$Z$ ($Z$=~P, As, Sb) crystallize in the C1$_b$ (space group $F\bar{4}3m$). The neutron diffraction measurements suggest a distortion of the structure which can promote the mobility. The possible modulation of the Lithium substructure will be a subject of further investigations. 
Optical reflectance measurements identified the direct band gap of these compounds. It has been found that LiMg$Z$ ($Z$=~P, As, Sb) are wide band gap semiconductors with band gaps ${E{_g}}$ of $\approx$ 2.3, 1.8 and 0.9 eV, respectively. The lineshape analysis of the $^7$Li static NMR spectra recorded as a function of temperature exhibit a clear tendency of decreasing the peak widths increasing the temperature, which is directly related with the increased Li ion mobility. 

\bigskip

\section{Acknowledgments}

Financial support by the Stiftung Rheinland-Pfalz for Innovation (Project 863) is gratefully acknowledged.

We are grateful to Dr. Wolfgang H. Meyer, Prof. Dr. Gerhard Wegner and Mr. Christoph Sieber from Max Planck Institute for Polymers Research for the DC conductivity measurements of the samples.

\end{document}